\documentclass[12pt]{iopart}
\usepackage{graphicx}
\usepackage{iopams}
\begin{document}

\title[Thermal Excitation of Multi-Photon Dressed States]{Thermal Excitation of Multi-Photon Dressed States\\ in Circuit Quantum Electrodynamics}

\author{J.~M.~Fink$^1$, M.~Baur$^1$, R.~Bianchetti$^1$, S.~Filipp$^1$, M.~G\"oppl$^1$, P.~J.~Leek$^1$, L.~Steffen$^1$, A.~Blais$^2$ and A.~Wallraff$^1$}

\address{$^1$ Department of Physics, ETH Zurich, CH-8093,
Zurich, Switzerland.}
\address{$^2$ D\'epartement de Physique, Universit\'e de
Sherbrooke, Qu\'ebec, J1K 2R1 Canada.}

\eads{\mailto{jfink@phys.ethz.ch}, \mailto{andreas.wallraff@phys.ethz.ch}}
\begin{abstract}

The exceptionally strong coupling realizable between superconducting qubits and photons stored in an on-chip microwave resonator allows for the detailed study of matter-light interactions in the realm of circuit quantum electrodynamics (QED). Here we investigate the resonant interaction between a single transmon-type multilevel artificial atom and weak thermal and coherent fields. We explore up to three photon dressed states of the coupled system in a linear response heterodyne transmission measurement. The results are in good quantitative agreement with a generalized Jaynes-Cummings model. Our data indicates that the role of thermal fields in resonant cavity QED can be studied in detail using superconducting circuits.
\end{abstract}

\pacs{42.50.Ct,42.50.Pq,03.67.Lx,85.35.Gv}

\section{Introduction}
In cavity quantum electrodynamics \cite{Raimond2001,Haroche2007,Walther2006,Ye2008} the fundamental interaction of matter and light is studied in a well controlled environment. On the level of individual quanta this interaction is governed by the Jaynes-Cummings Hamiltonian \cite{Jaynes1963}. In early experiments the quantization of the electromagnetic field was observed in cavity QED with Rydberg atoms by measurements of collapse and revival  \cite{Rempe1987}. Similarly, the $\sqrt{n}$ scaling of the atom/photon coupling strength with the number of photons $n$ has been observed in the time domain by measuring $n$ photon Rabi oscillations using coherent states \cite{Brune1996} and Fock states \cite{Varcoe2000,Bertet2002}. In the frequency domain the $\sqrt{n}$ scaling can be extracted from spectroscopic cavity transmission measurements. Initial attempts employing pump and probe spectroscopy with alkali atoms \cite{Thompson1998} were inconclusive. In a recent experiment however the two-photon vacuum Rabi resonance was resolved using high power nonlinear spectroscopy \cite{Schuster2008}. At the same time,  we observed the quantum nonlinearity by measuring the spectrum of two photons and one artificial atom in circuit QED \cite{Fink2008}. In these measurements the originally proposed pump and probe spectroscopy scheme \cite{Thompson1998}
was used. Similarly the $n$ photon Rabi mode splitting was studied using multi-photon transitions up to $n=5$ \cite{Bishop2009a}. In the dispersive regime the photon number splitting of spectroscopic lines \cite{Schuster2007a} provides similar evidence for the quantization of microwave radiation in circuit QED. Further experimental progress in the time domain has enabled the preparation and detection of coherent states \cite{Johansson2006}, photon number states \cite{Hofheinz2008} and arbitrary superpositions of photon number states \cite{Hofheinz2009} in circuits. Using Rydberg atoms quantum jumps of light have been observed \cite{Gleyzes2007,Guerlin2007} and Wigner functions of Fock states have been reconstructed \cite{Deleglise2008}. These experiments demonstrate the quantum nature of light by measuring the nonlinear $\sqrt{n}$ scaling of the dipole coupling strength with the discrete number of photons $n$ in the Jaynes-Cummings model \cite{Carmichael1996}.

In this letter we extend our earlier spectroscopic study of the $\sqrt{n}$ scaling \cite{Fink2008} in the circuit QED architecture \cite{Blais2004,Wallraff2004b}. We present a measurement of the photon/atom energy spectrum up to $n=3$ excitations using both thermal and coherent fields. A solid state qubit with a large effective dipole moment plays the role of the atom. It is coupled to weak microwave radiation fields in a coplanar waveguide resonator which provides a very large electric field strength per photon. The coupling strength of a few hundred MHz \cite{Schoelkopf2008}, much larger than in implementations with natural atoms, can freely be chosen over a wide range of values and remains fixed in time. The qubit level spectrum is in situ tunable by applying flux to the circuit and the internal degrees of freedom can be prepared and manipulated with microwave fields.

Similar solid state circuit implementations of cavity QED have enabled a remarkable number of novel quantum optics \cite{Wallraff2004b,Astafiev2007,Houck2007,Schuster2007a,Fink2008,Hofheinz2008,Fragner2008, Deppe2008,Bishop2009a,Hofheinz2009,Baur2009,Fink2009} and quantum computation \cite{Majer2007,Sillanpaa2007,DiCarlo2009} experiments.

\section{Sample and experimental setup}
\begin{figure}[t]\center
\includegraphics[width=1 \columnwidth]{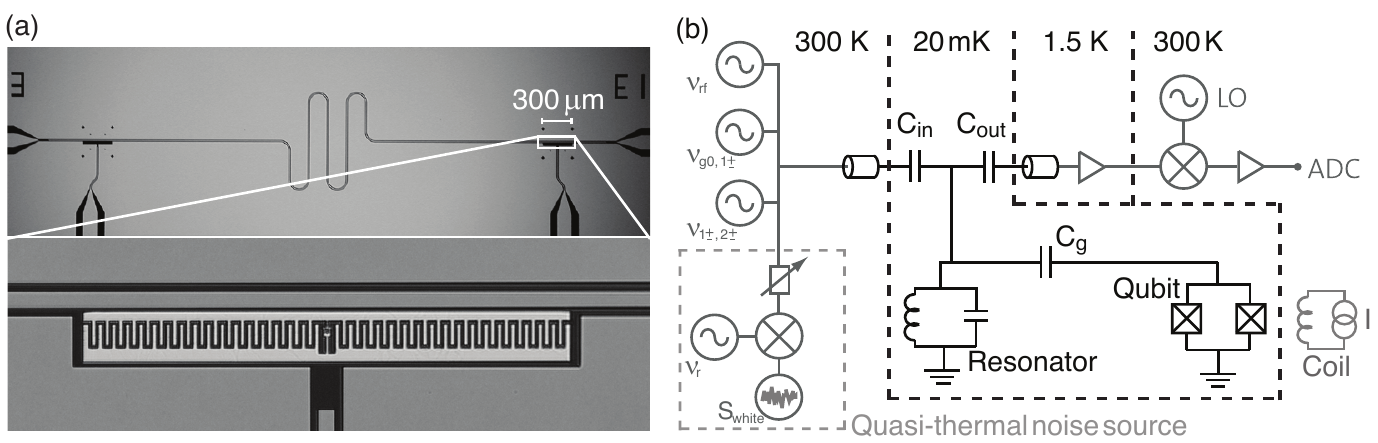}
\footnotesize \caption{Sample and experimental setup. (a)~Optical images of the superconducting coplanar waveguide resonator (top) with the transmon type superconducting qubit embedded at the position indicated. The qubit of dimensions $300\times30~\mu\textrm{m}^2$ is shown on the bottom. (b)~Simplified circuit diagram of the experimental setup. The qubit is capacitively coupled to the resonator through $C_{g}$ and flux tuned with a current biased ($I$) external miniature superconducting coil. The resonator, represented by a parallel LC circuit, is coupled to input and output transmission lines via the capacitors $C_\textrm{in}$ and $C_\textrm{out}$. A quasi thermal field can be generated by modulating a microwave carrier tone $\nu_r$ with a large bandwidth pseudo-random noise spectrum $S_\textrm{white}$. The noise signal is filtered, attenuated and added to pump ($\nu_{g0,1\pm}$, $\nu_{1\pm,2\pm}$) and probe ($\nu_{\textrm{rf}}$) tones. The resulting microwave fields are further attenuated and applied to the resonator input. After amplification with an ultra low noise amplifier the transmitted probe signal $\nu_{\textrm{rf}}$ is downconverted with a local oscillator (LO) and digitized with an analog to digital converter (ADC).} \label{fig1}
\end{figure}

For the experiments presented here we use a transmon type qubit \cite{Koch2007,Schreier2008}, which is a charge-insensitive superconducting qubit design derived from the Cooper pair box \cite{Bouchiat1998}, as the artificial atom. Its transition frequency is given by $\omega_{g,e}/2\pi\simeq\sqrt{8 E_{C} E_{J}(\Phi)}-E_{C}$ with the single electron charging energy $E_{C}/h \approx0.232 \, ~\textrm{GHz}$, the flux controlled Josephson energy $E_{J}(\Phi)=E_{{J,max}}|\cos{(\pi \Phi/\Phi_{0})}|$ and $E_{{J,max}}/h \approx 35.1 \, \textrm{GHz}$, as determined by spectroscopic measurements. The two characteristic energies  $E_{J}$ and $E_{C}$ define the full level spectrum of the qubit where the eigenenergy of level $l$ is approximately given as $E_l\simeq-E_J+\sqrt{8 E_C E_J} (l+1/2) - E_C/12 (6l^2+6l+3)$ \cite{Koch2007}. The cavity is realized as a coplanar resonator with bare resonance frequency $\nu_{r}\approx6.44 \, \textrm{GHz}$ and photon decay rate $\kappa/2\pi\approx1.6 \, \textrm{MHz}$. Details of the resonator design and fabrication can be found in Ref.~\cite{Goppl2008}. Optical microscope images of the sample are shown in Fig.~\ref{fig1}a. A simplified electrical circuit diagram of the setup is shown in Fig.~\ref{fig1}b.

\section{Generalized Jaynes-Cummings model}
\begin{figure}[t]\center
\includegraphics[width=0.45 \columnwidth]{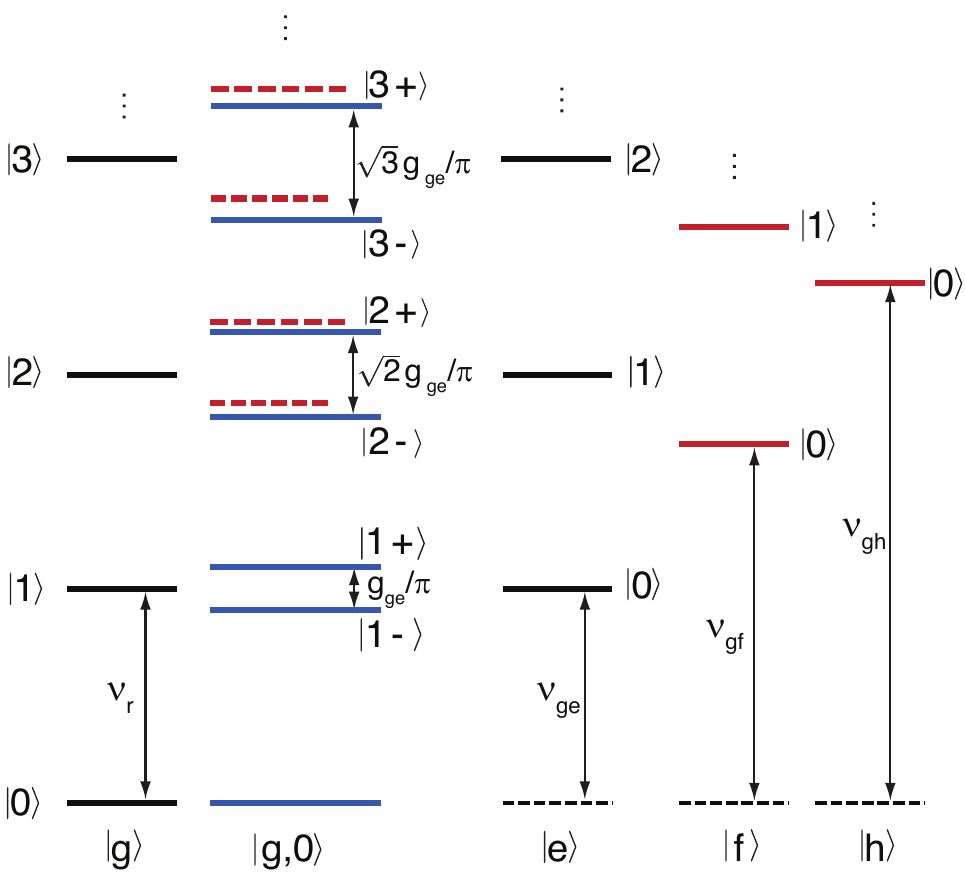}
\footnotesize \caption{Sketch of the energy level diagram of a resonant ($\nu_{r}=\nu_{g,e}$) cavity QED system. The uncoupled qubit states $|g\rangle, |e\rangle$, $|f\rangle$ and $|h\rangle$ (from left to right) and the photon states $|0\rangle$, $|1\rangle$, $|2\rangle$, $|3\rangle$, ... (from bottom to top) are shown with black and red solid lines. The dipole coupled dressed states are shown in blue and the shifts due to the $|f,0\rangle$ and $|h,0\rangle$ levels are indicated by red dashed lines.} \label{fig2}
\end{figure}

The transmon qubit is a superconducting circuit with a nonlinear energy level spectrum. In many experiments the nonlinearity, which can be adjusted by circuit design and fabrication, is sufficient to correctly model it as a two level system. For the experimental results presented in this work it is however essential to treat the qubit as a multilevel system taking into account the coupling of all relevant transmon levels to the cavity photons. The physics of this multilevel artificial atom strongly coupled to a single mode of the electromagnetic field is described by a generalized Jaynes-Cummings model with the Hamiltonian
\begin{equation}\label{jch}
\hat{\mathcal{H}}= \hbar\omega_{r}\, \hat{a}^\dagger \hat{a}+ \sum_{l=e,f,h,...}\left( \hbar\omega_{l}\,\hat{\sigma}_{l,l}\, +\hbar g_{l-1,l}\,(\hat{\sigma}^\dagger_{l-1,l}\,\hat{a} +\hat{a}^\dagger\hat{\sigma}_{l-1,l}\,)\right) \, .
\end{equation}
Here, $\hbar\omega_{{l}}$ is the energy of the $l$'th excited state $|l\rangle$ of the multilevel artificial atom, $\omega_{r}$ is the frequency of the resonator field and $g_{l-1,l}$ is the coupling strength of the transition $l-1 \rightarrow l$ and one photon. $\hat{a}^{\dagger}$ and $\hat{a}$ are the raising and lowering operators acting on the field with photon number $n$ and $\hat{\sigma}_{i,j}=|i\rangle\langle j|$ are the corresponding operators acting on the qubit states. In Fig.~\ref{fig2}, a sketch of the energy level diagram of the resonantly coupled qubit-resonator system ($\nu_{r}=\nu_{g,e}$) is shown for up to three photons $n=0,1,2,3$ and the first four transmon levels $l=g,e,f$ and $h$.

Considering the first two levels of the artificial atom at zero detuning ($\Delta\equiv\nu_{r}-\nu_{g,e}=0$) the eigenstates of the coupled system are the symmetric $(|g,n\rangle+|e,n-1\rangle)/\sqrt{2}\equiv|n+\rangle$ and antisymmetric $(|g,n\rangle-|e,n-1\rangle)/\sqrt{2}\equiv|n-\rangle$ qubit-photon superposition states, see Fig.~\ref{fig2}. For $n=1$, the coupled one photon one atom eigenstates are split due to the dipole interaction \cite{Agarwal1984,Sanchez1983}.
These states have been observed in hallmark experiments demonstrating the strong coupling of a single atom and a single photon both spectroscopically as a vacuum Rabi mode splitting with on average a single atom \cite{Thompson1992} and also without averaging using alkali atoms \cite{Boca2004}, in superconducting circuits \cite{Wallraff2004b} and semiconducting quantum dot systems \cite{Reithmaier2004,Yoshie2004a}, or in time resolved measurements as vacuum Rabi oscillations at frequency $2g_{g,e}$ with Rydberg atoms \cite{Brune1996,Varcoe2000} and superconducting circuits \cite{Johansson2006,Hofheinz2008}. At least in principle however the observation of the first doublet $|1\pm\rangle$ alone, could also be explained as normal mode splitting in a classical linear dispersion theory \cite{Zhu1990}.
The Jaynes-Cummings model however predicts a characteristic nonlinear scaling of this frequency as $\sqrt{n}\, 2 g_{g,e}$ with the number of excitations $n$ in the system.
In the general multilevel case, the higher energy atomic levels renormalize the dipole coupled dressed state energies. This causes frequency shifts in the excitation spectrum as indicated in Fig.~\ref{fig2} and the simple $\sqrt{n}$ scaling is slightly modified.

\section{Vacuum Rabi mode splitting in the presence of a thermal field}
In this section we investigate vacuum Rabi resonances both in the presence of a weak thermal background field and also in the presence of externally applied quasi-thermal fields. To our knowledge this is the first experiment where thermally excited multi-photon dressed states are studied systematically in a cavity QED system.

\begin{figure}[b]\center
\includegraphics[width=0.9 \columnwidth]{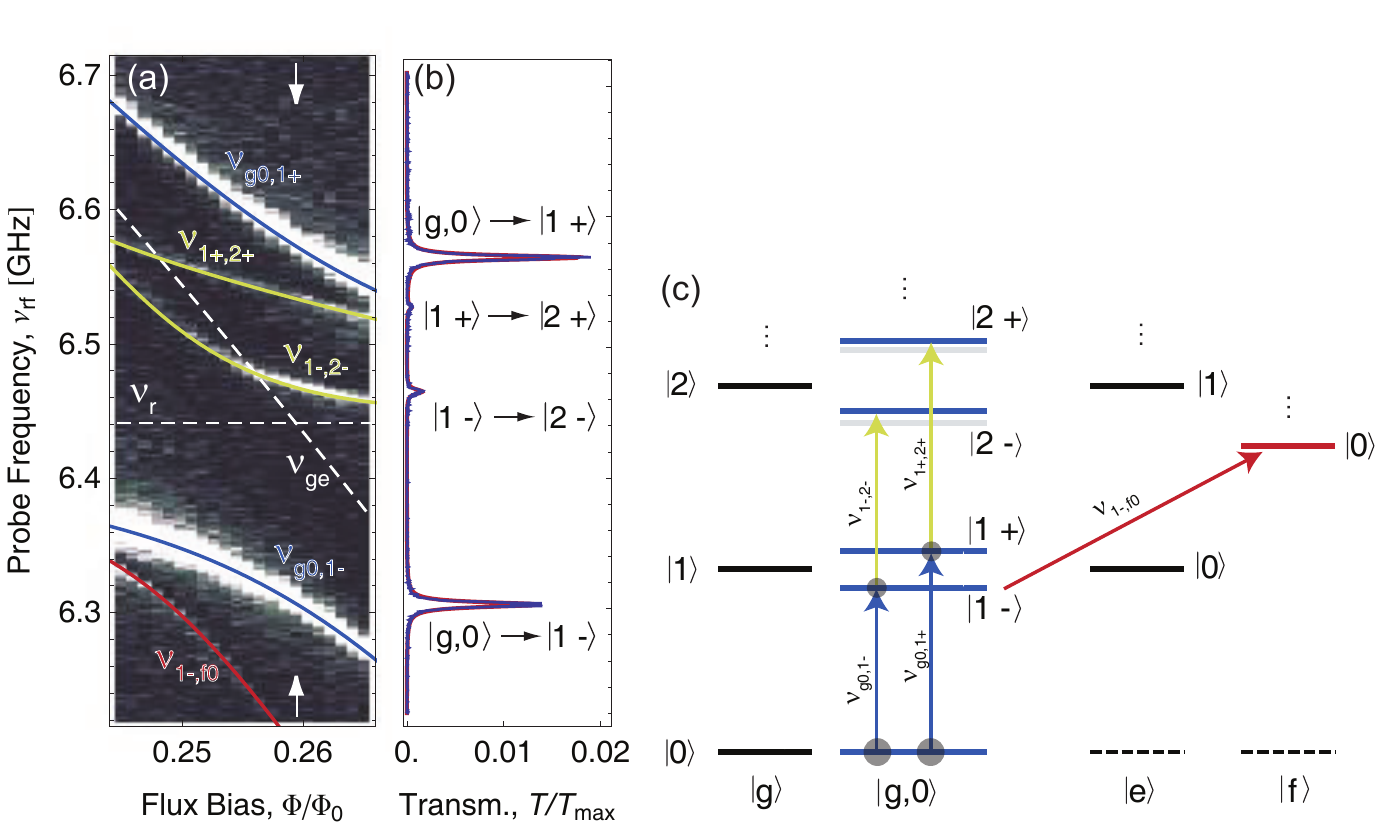}
\footnotesize \caption{Vacuum Rabi mode splitting with a weak coherent probe tone. (a)~Measured resonator transmission spectra versus external flux $\Phi$ close to degeneracy. Black indicates low and white high transmission $T$. The solid lines show dressed state energies as obtained numerically and the dashed lines indicate the bare resonator frequency $\nu_{r}$ and the qubit transition frequency $\nu_{\textit{g,e}}$. (b)~Resonator transmission $T$ at degeneracy as indicated with arrows in panel (a), with a line fit to four Lorentzians in red. (c)~Corresponding energy level spectrum (similar to Fig.~\ref{fig2}) with observed transitions indicated by arrows and small thermal population indicated with gray circles.} \label{fig3}
\end{figure}

\begin{figure}[b]\center
\includegraphics[width=0.9 \columnwidth]{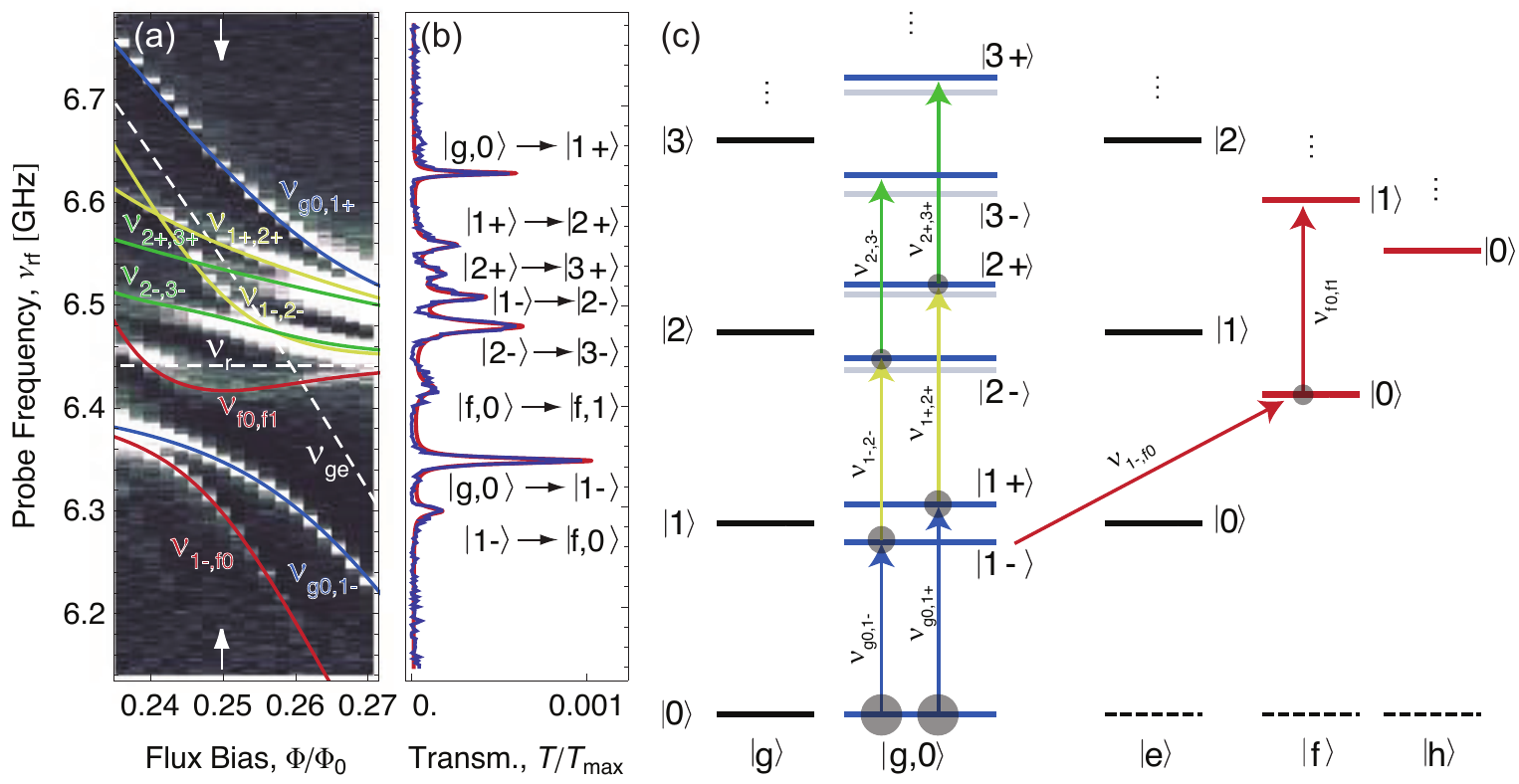}
\footnotesize \caption{Vacuum Rabi mode splitting in the presence of a thermal field. (a)~Cavity transmission $T$ as in Fig.~\ref{fig3} with an additional quasi-thermal field of temperature $T_r \sim 0.4$ \textrm{K} applied to the resonator input populating the $|1\pm\rangle$, $|2\pm\rangle$ and $|f,0\rangle$ states. (b)~Transmission spectrum at $\Phi/\Phi_0=0.25$, indicated by arrows in (a). (c)~Corresponding energy level spectrum (similar to Fig.~\ref{fig2}) with observed transitions indicated by arrows and induced thermal population indicated with gray circles.} \label{fig5}
\end{figure}

Experimentally, the coupled circuit QED system is prepared in its ground state $|g,0\rangle$ by cooling it to temperatures below $<20\, \textrm{mK}$ in a dilution refrigerator. Measuring the cavity transmission spectrum $T$ in the anti-crossing region of the qubit transition frequency $\nu_{g,e}$ and the cavity frequency $\nu_r$
yields transmission maxima at frequencies corresponding to transitions from $|g,0\rangle$ to the first doublet $|1\pm\rangle$ of the Jaynes-Cummings ladder, see Fig.~\ref{fig3}a. On resonance, see Fig.~\ref{fig3}b, we extract the coupling strength of $g_{{g,e}}/2\pi = 133 \, \rm{MHz}$. This is a clear indication that the strong coupling limit $g_{g,e} \gg \kappa, \gamma$, with photon decay and qubit decoherence rates $\kappa/2\pi, \gamma/2\pi \sim 1\, \rm{MHz}$, is realized. Solid lines in Fig.~\ref{fig3}a (and Fig.~\ref{fig5}a) are numerically calculated dressed state frequencies, see Fig.~\ref{fig3}c, with the qubit and resonator parameters as stated above. For the calculation, the qubit Hamiltonian is diagonalized exactly in the charge basis. The qubit states $|g\rangle$ and $|e\rangle$ and the flux dependent coupling constant $g_{g,e}$ are then incorporated in the Jaynes-Cummings Hamiltonian Eq.~(\ref{jch}). Its numerical diagonalization yields the dressed states of the coupled system without any fit parameters.

In addition to the expected spectral lines corresponding to the transition from the ground state $|g,0\rangle$ to the first doublet states $|1\pm\rangle$, we observe three lines with very low intensities, see Figs.~\ref{fig3}a and b. These additional transitions are visible because the system is excited by a small thermal background field with a cavity photon number distribution given by the Bose-Einstein distribution. This thermal field is a consequence of incomplete thermalization of the room temperature black-body radiation at the input and output ports of the resonator. A quantitative analysis taking into account the two photon states $|2\pm\rangle$ and the presence of the higher energy qubit levels $f$ and $h$ in Eq.~(\ref{jch}) yields the transition frequencies indicated by yellow and red solid lines in Fig.~\ref{fig3}. For this analysis the coupling constants $g_{e,f}=184$ \textrm{MHz} and $g_{f,h}=221$ \textrm{MHz} of higher energy qubit levels to the cavity mode, obtained from exact diagonalization of the qubit Hamiltonian, have been included. We thus identify two of the additional spectral lines as transitions between the first $|1\pm\rangle$ and second $|2\pm\rangle$ doublet states of the resonant Jaynes Cummings ladder, see Fig.~\ref{fig3}c. The lowest frequency additional spectral line corresponds to a transition from the antisymmetric doublet state with one photon $|1-\rangle$ to the qubit $f$ level without a photon $|f,0\rangle$. The details of the thermally excited transmission spectrum can be used as a sensitive probe for the cavity field temperature. Analyzing the amplitudes of the Rabi splitting spectrum with a quantitative master equation model \cite{Bishop2009a}, leads to an estimated photon temperature of $T_r\simeq 0.2$ \textrm{K} which corresponds to a relatively high mean thermal occupation number of $\bar{n}_{th}\simeq 0.3$ photons for the data presented in Fig.~3a. Careful filtering and thermalization at the input and output ports results in a typical cavity photon temperature of $ < 90\, \textrm{mK}$ and $ < 54\, \textrm{mK}$ ($\bar{n}_{th} < 0.03$ and $\bar{n}_{th} < 0.003$) as reported in Refs.~\cite{Fragner2008} and \cite{Bishop2009a}.

In order to access the three photon doublet states $|3\pm\rangle$ of the coupled multi-photon multilevel-atom system we use externally applied broadband quasi-thermal fields. In this new approach the dressed eigenstates are populated according to a thermal distribution depending on the chosen thermal field temperature. This allows to investigate the flux dependence of all resolvable spectral lines for a given effective resonator mode temperature $T_r$ in a single experimental run. The spectrum of a one dimensional black body such as the considered cavity is given as $S_{1D}(\nu)=h\nu/[\exp{(h\nu/k_B T})-1]$. At a temperature of $T_r > h \nu_r/k_B\simeq300$ \textrm{mK} this energy spectrum is flat with a deviation of $<5\%$ within a 500 \textrm{MHz} band centered at the cavity frequency $\nu_r\simeq6.44$ \textrm{GHz}. It is therefore a very good approximation to make use of a white noise spectrum in the narrow frequency band of the experimentally investigated transition frequencies centered at $\nu_r$, in order to generate a quasi-thermal field of temperatures $T_r>300$ \textrm{mK} and populate the considered cavity mode with thermal photons.

In order to generate such a spectrum a carrier microwave tone at frequency $\nu_r$ is modulated with a low frequency large bandwidth quasi-random noise spectrum $S_\textrm{white}$ using a mixer, see Fig.~\ref{fig1}b. This approximately yields a microwave frequency white noise spectrum with a bandwidth of $500$ \textrm{MHz} centered symmetrically around the cavity frequency $\nu_r$. Using tunable attenuators, we can adjust the noise power spectral density over a wide range of values. For this experiment we adjust it such that the thermal population of the cavity mode is on the order of $\bar{n}_{th} \sim 0.9$ corresponding to a temperature of $T_r \sim 0.4$ \textrm{K}. This noise spectrum constitutes a sufficient approximation of a black body thermal noise source for the considered 1D cavity mode, temperature and frequency. At the same time, the chosen mean thermal population $\bar{n}_{th}\sim 0.9$ allows to observe all allowed transitions between the ground state $|g\rangle$ and the three photon doublet states $|3\pm\rangle$.

In the presence of the thermal field, we probe the cavity transmission spectrum as a function of flux in the anticrossing region, see Fig.~\ref{fig5}a, with a weak coherent probe tone $\nu_\textrm{rf}$ in the linear response limit. In this limit the weak probe tone is only a small perturbation to the field and no multi-photon transitions are induced. In this measurement we resolve all allowed transitions between the thermally occupied dipole coupled states in the generalized Jaynes-Cummings ladder. The solid lines are again the calculated dressed state transition energies which agree well with the observed spectral lines. In Fig.~\ref{fig5}b, a cavity transmission measurement at flux $\Phi/\Phi_0=0.25$ is shown. We identify 8 allowed transitions, compare with Fig.~\ref{fig5}c. It follows that the states $|1\pm\rangle$, $|2\pm\rangle$ and also $|f,0\rangle$ are thermally populated.

\begin{figure}[b]\center
\includegraphics[width=0.35 \columnwidth]{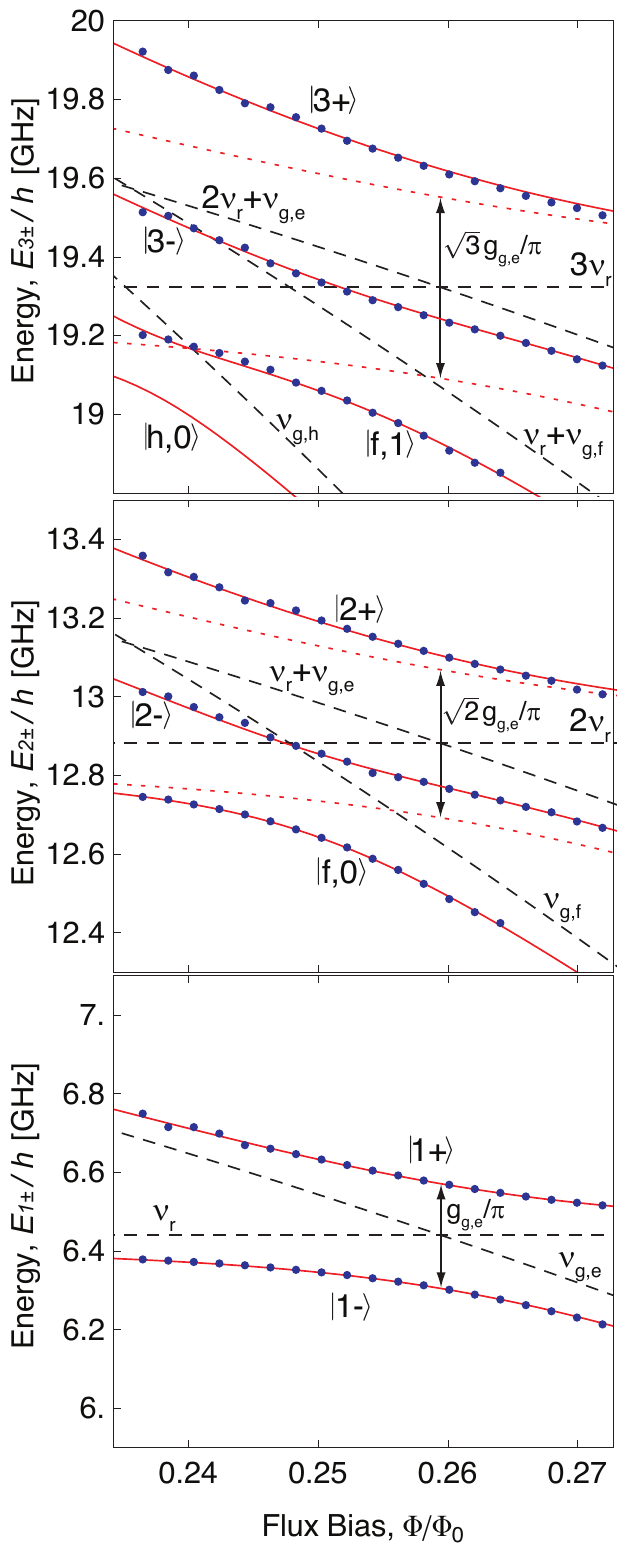}
\footnotesize \caption{Measured energy level diagram of the three-photon / artificial atom system. Measured dressed state energies (blue dots) reconstructed from extracted transition frequencies from data in Fig.~\ref{fig5} compared to calculated uncoupled cavity and qubit levels (dashed black lines), the calculated dressed state energies in the qubit two-level approximation (dotted red lines) and to the corresponding calculation including four qubit levels (solid red lines).} \label{fig6}
\end{figure}

In the two-level-atom approximation, transitions between symmetric and antisymmetric doublet states are forbidden at degeneracy. In the generalized Jaynes-Cummings model the dressed state transition matrix elements are renormalized due to higher qubit levels. Numerical diagonalization however shows that the matrix elements squared, which are related to the amplitude of the expected spectral lines, are 140 (6) times smaller for the symmetry changing transitions $|1-\rangle\rightarrow|2+\rangle$ ($|1+\rangle\rightarrow|2-\rangle$) than for the observed symmetry preserving transitions $|1-\rangle\rightarrow|2-\rangle$ ($|1+\rangle\rightarrow|2+\rangle$) at degeneracy. Similarly, for the transitions $|2-\rangle\rightarrow|3+\rangle$ ($|2+\rangle\rightarrow|3-\rangle$) the matrix elements squared are 235 (16) times smaller than the measured transitions $|2-\rangle\rightarrow|3-\rangle$ ($|2+\rangle\rightarrow|3+\rangle$) at degeneracy. Therefore transitions between symmetric and antisymmetric doublet states are not resolved in our experiment. Symmetry changing transitions populating the antisymmetric states $|2-\rangle$ and $|3-\rangle$ have larger matrix elements than symmetry changing transitions populating the symmetric states $|2+\rangle$ and $|3+\rangle$ because the former are closer in frequency to the qubit levels $|f,0\rangle$ and $|h,0\rangle$. Similarly, the matrix element for the transition $|1-\rangle\rightarrow|f,0\rangle$ is 34 times larger than for the transition $|1+\rangle\rightarrow|f,0\rangle$ at degeneracy. The latter is therefore also not observed in the experimental data. In addition to the transition $|1-\rangle\rightarrow|f,0\rangle$, also seen in the data presented in Fig.~\ref{fig3}, we observe a transmission line which corresponds to the transition $|f,0\rangle\rightarrow|f,1\rangle$, see Fig.~\ref{fig5}b.
A numerical calculation shows that the matrix element is 5 times larger at degeneracy than $|2-\rangle\rightarrow|f,1\rangle$ and 7 times larger than $|f,0\rangle\rightarrow|h,0\rangle$ which in principle could also have been observed. All transitions observed in the experimental data are in qualitative agreement with the calculated matrix elements stated above.

In Fig.~\ref{fig6} the complete measured level spectrum of the bound photon/atom system up to the third excitation is shown. To calculate the absolute energies of the levels (blue dots) we extract the transition frequencies from data presented in Fig.~\ref{fig5} with Lorentzian line fits and add them accordingly. For the first doublet states $|1\pm\rangle$ we find excellent agreement with both a simple two-level atom Jaynes-Cummings model (dotted red lines) as well as the generalized multilevel Jaynes-Cummings model (solid red lines). In the case of the second $|2\pm\rangle$ and third doublet states $|3\pm\rangle$ we find considerable frequency shifts with regard to the two level model (compare dotted and solid red lines) but excellent agreement with the generalized model taking account the additional qubit levels. Furthermore it can be seen in Fig.~\ref{fig6} that the negative anharmonicity of the transmon qubit, together with the strong dipole coupling, causes large frequency shifts of the antisymmetric dressed levels $|2-\rangle$ and $|3-\rangle$ since they are closer in frequency to the qubit levels $|f,0\rangle$ and $|f,1\rangle$, $|h,0\rangle$. This leads to a small reduction of the $\sqrt{n}$ nonlinearity which is in agreement with the numerical results.

\section{Vacuum Rabi mode splitting with two pump and one probe tone}
\begin{figure}[t]\center
\includegraphics[width=1 \columnwidth]{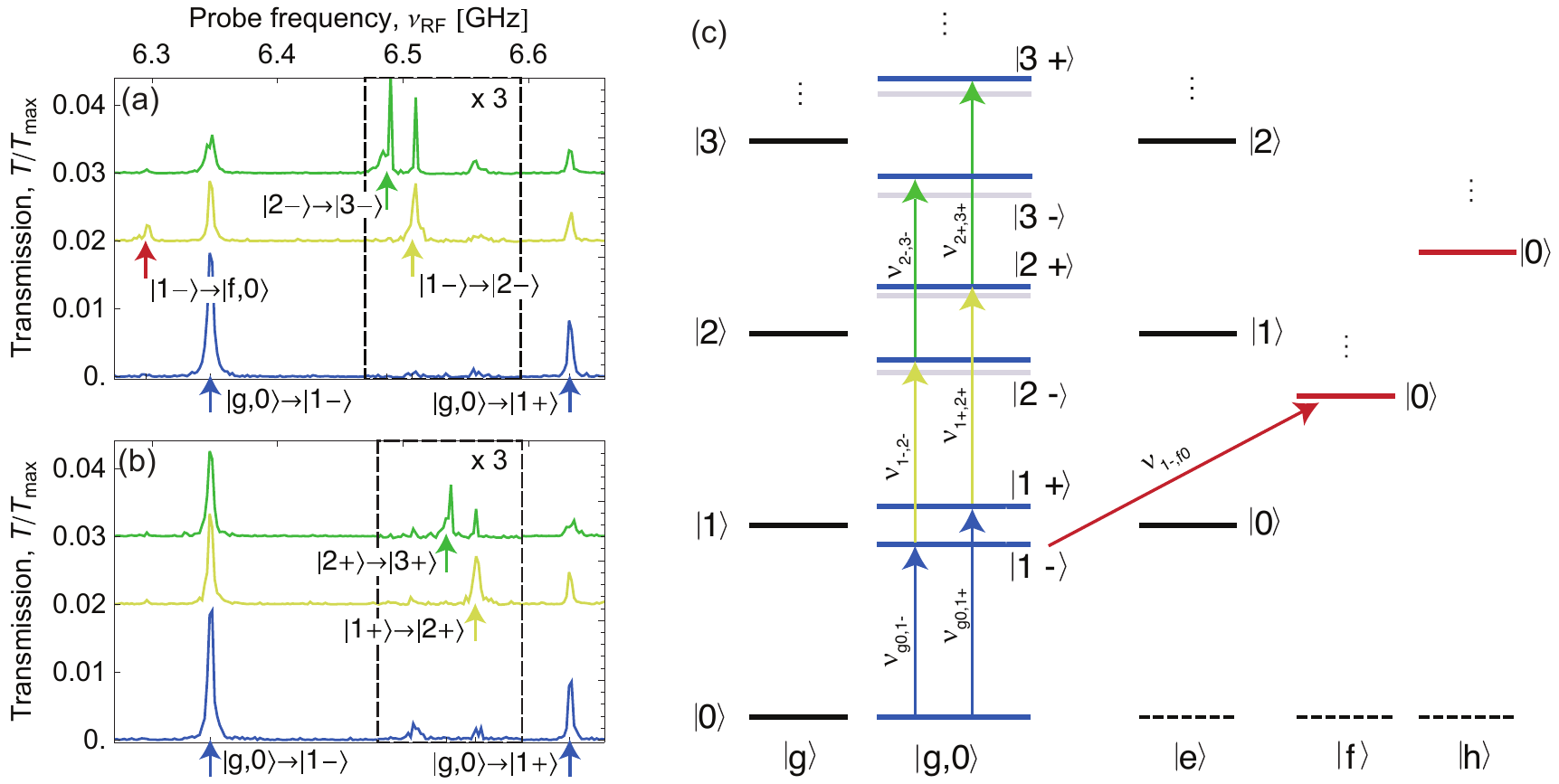}
\footnotesize \caption{Vacuum Rabi mode splitting with one probe tone and zero, one or two coherent pump tones at flux $\Phi/\Phi_0\approx0.25$ close to degeneracy. (a)~Measured resonator transmission spectra $T/T_{max}$ without a pump tone (blue) with one pump tone $\nu_{g0,1-}$ (yellow) and with two pump tones $\nu_{g0,1-}$ and $\nu_{1-,2-}$ (green) applied. The spectra are offset by 0, 0.02 and 0.03 $T/T_\textrm{max}$ and the boxed area is scaled in amplitude by a factor of 3 for better visibility. Vertical arrows indicate numerically calculated transition frequencies. (b)~Similar measurement of resonator transmission $T$ for the case of no (blue), $|1+\rangle$ (yellow) and both $|1+\rangle$ and $|2+\rangle$ symmetric dressed states pumped coherently. (c)~Energy level spectrum (similar to Fig.~\ref{fig2}) with relevant transitions indicated by arrows.} \label{fig4}
\end{figure}

In order to probe the excitation spectrum of the two and three photon doublet states we can also follow the pump and probe spectroscopy scheme similar to the one presented in Refs.~\cite{Thompson1998,Fink2008}, where one of the first doublet states $|1\pm\rangle$ is coherently pumped and the transition to the states $|2\pm\rangle$ is probed. This technique avoids large intra-cavity photon numbers which are needed in high-drive and elevated temperature experiments. In analogy to the previous section we wait for the system to equilibrate with the cold environment to prepare the ground state $|g,0\rangle$. The qubit is then tuned close to degeneracy where $\Phi/\Phi_0\approx0.25$. We weakly probe the resonator transmission spectrum as shown in Fig.~\ref{fig4} (blue lines). In a second step we apply a pump tone at frequency $\nu_{g0,1-}$ ($\nu_{g0,1+}$) occupying the dressed state $|1-\rangle$ ($|1+\rangle$) and probe the system again, see Fig.~\ref{fig4}a (b) yellow line. Clearly the transitions $|1\pm\rangle\rightarrow|2\pm\rangle$ become visible at the calculated eigenenergy which is indicated with yellow vertical arrows in Figs.~\ref{fig4}a, b and c. We also note that the state $|f,0\rangle$ is populated by the probe tone via the transition $|1-\rangle\rightarrow|f,0\rangle$, see red arrows in Figs.~\ref{fig4}a and c. In a last step we apply two pump tones at frequencies $\nu_{g0,1-}$ and $\nu_{1-,2-}$, see Fig.~\ref{fig4} a (green line), or at frequencies $\nu_{g0,1+}$ and $\nu_{1+,2+}$, see Fig.~\ref{fig4} b (green line) respectively. The three-photon one-qubit dressed state transitions $|2\pm\rangle\rightarrow|3\pm\rangle$ become visible in the spectrum, see green vertical arrows. At the same time, transitions from the ground state are found to saturate considerably when the pump tones are turned on, compare the amplitudes of the spectral lines at the frequency indicated by the left blue arrow in Fig.~\ref{fig4} a, or similarly by the right blue arrow in figure Fig.~\ref{fig4} b. This is expected since the occupation probability of the ground state is reduced and the transition starts to become saturated when the pump tones are turned on.

Again, the observed transition frequencies are in good agreement with the calculated dressed state transition energies indicated by vertical arrows in Figs.~\ref{fig4}a and b. Similar to the experiment described in the last section, additional spectral lines with low intensity, see Fig.~\ref{fig4}a and b blue lines, occur because of a small probability of occupation of the first doublet due to the residual thermal field. In comparison to the data shown in Fig.~\ref{fig5} no transition $|f,0\rangle\rightarrow|f,1\rangle$ is observed because the level $|f,0\rangle$ is neither thermally, nor coherently populated here.

\section{Summary}
We extended our previous work \cite{Fink2008} by introducing thermal fields to populate the dressed eigenstates in a resonant cavity QED system. In addition to the one and two photon/atom superposition states we report a measurement of the three photon doublet using both thermal and coherent fields. The results are in good agreement with a generalized multilevel-atom Jaynes-Cummings Hamiltonian without any fit parameters. Related results have been reported in an atomic system \cite{Schuster2008} and in circuit QED \cite{Bishop2009a}. Similar effects may be interesting to approach also in semiconducting cavity QED systems. It has been shown that cavity QED with superconducting circuits can be a sensitive probe for thermal fields. A more detailed quantitative analysis of the thermally excited vacuum Rabi spectra could be of interest in the context of environmentally induced dissipation and decoherence, thermal field sensing and the cross-over from the quantum to the classical regime of cavity QED.

\section{Acknowledgements}
This work was supported by EuroSQIP, SNF Grant No. 200021-111899 and ETHZ. A.~B.~was supported by NSERC, CIFAR and the Alfred P.~Sloan Foundation.

\section*{References}
\bibliographystyle{unsrt}
\bibliography{QudevRefDB}

\end{document}